\documentclass[sigconf,nonacm]{acmart}
\usepackage{xspace}
\usepackage{array}
\usepackage{multirow}

\AtBeginDocument{%
  }

\newcommand{\sysname}{{PoVSmith}\xspace}
\newcommand{\sysnameg}{{PoVSmith$_G$}\xspace}
\newcommand{\sysnamem}{{PoVSmith$_M$}\xspace}
\newcommand{\tool}{\sysname}
\newcommand{\totalCallPath}{216\xspace}
\newcommand{\totalCorrectCallPath}{207\xspace}
\newcommand{\totalentry}{158\xspace}
\newcommand{\totaltest}{152\xspace}
\newcommand{\totalcompile}
{141\xspace}

\newcommand{\totalsuccess}{84\xspace}
\newcommand{\successrate}{55\%\xspace}
\newcommand{\totalcorrectlabel}{103\xspace}
\newcommand{\totalApps}{33\xspace}
\newcommand{\numOfCve}{20\xspace}
\newcommand{\codefont}[1]{\footnotesize{\texttt{#1}}\normalsize}

\newcolumntype{L}[1]{>{\raggedright\let\newline\\\arraybackslash\hspace{0pt}}m{#1}}
\newcolumntype{C}[1]{>{\centering\let\newline\\\arraybackslash\hspace{0pt}}m{#1}}
\newcolumntype{R}[1]{>{\raggedleft\let\newline\\\arraybackslash\hspace{0pt}}m{#1}}

\begin{document}
\settopmatter{printfolios=true}

\title{Generating Proof-of-Vulnerability Tests to Help Enhance the Security of Complex Software}

\author{Shravya Kanchi}
\affiliation{
  \institution{Virginia Tech}
  \city{Blacksburg}
  \state{Virginia}
  \country{USA}
}

\author{Xiaoyan Zang}
\affiliation{
  \institution{Virginia Tech}
  \city{Blacksburg}
  \state{Virginia}
  \country{USA}
}

\author{Ying Zhang}
\affiliation{
  \institution{Virginia Tech}
  \city{Blacksburg}
  \state{Virginia}
  \country{USA}
}

\author{Danfeng (Daphne) Yao}
\affiliation{
  \institution{Virginia Tech}
  \city{Blacksburg}
  \state{Virginia}
  \country{USA}
}

\author{Na Meng}
\affiliation{
  \institution{Virginia Tech}
  \city{Blacksburg}
  \state{Virginia}
  \country{USA}
}

\begin{abstract}
Developers create modern software applications (Apps) on top of third-party libraries (Libs). 
When library vulnerabilities are reachable through application code, the applications can be vulnerable to software supply chain attacks. 
Prior work shows that developers often require concrete and executable evidence, i.e., \textbf{proof-of-vulnerability (PoV) tests}, to decide whether a reported dependency vulnerability poses a practical security risk to their application.
However, manually crafting such tests is challenging, and existing tool support is insufficient to automate the procedure.

To streamline test generation,
we created \sysname---a new approach that combines call path analysis, exemplar test, code context, and feedback into multiple prompts to guide a coding agent (i.e., Codex) and a large language model (i.e., GPT)  for test generation, execution, and assessment. 
We evaluated \sysname on \totalApps $\langle App, Lib\rangle$ Java program pairs, where each App depends on a vulnerable Lib. 
\tool revealed \totalentry unique application-level entry points (i.e., public methods) calling vulnerable library APIs; \totaltest (96\%) of them were correctly found, together with the call paths properly recognized. With such method call information,
\tool generated \totaltest tests, \totalsuccess (\successrate) of which demonstrated feasible ways of attacking Apps by exploiting Lib vulnerabilities.
\tool substantially outperforms the state-of-the-art LLM-based approach, as it reduces human involvement while dramatically improving test quality. 

Our work contributes (1) a novel approach of agent-based test generation, (2) an
iterative code refinement process driven by execution feedback, and (3) 
LLM-based quality assessment grounded in both the test context and execution logs.
\end{abstract}



\maketitle

\section{Introduction}

\begin{figure}[h]
    \centering
    \vspace{-1em}
    \includegraphics[width=\linewidth]{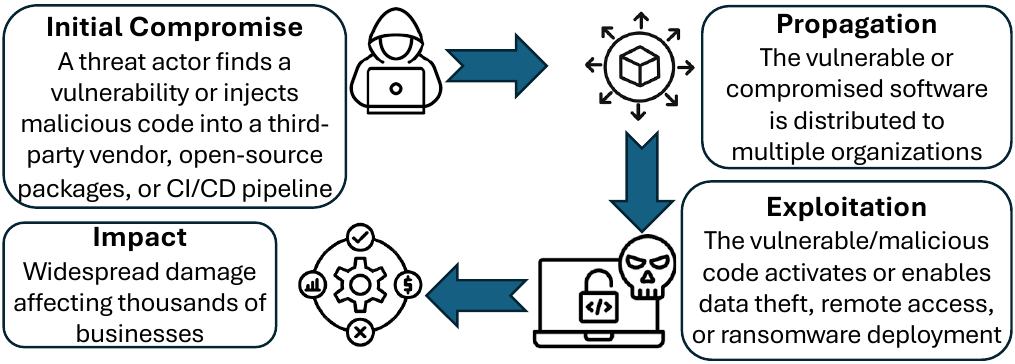}
    \vspace{-2em}
    \caption{The threat model of software supply chain attacks}
    \label{fig:threat-model}
    \vspace{-1em}
\end{figure}

\textbf{Software supply chain (SSC)} is an aggregation of all the people, processes, and technologies involved in producing or updating a piece of software~\cite{supply-chain-def}.
\textbf{Software supply chain vulnerabilities} are security flaws, in external dependencies or third-party libraries (Libs) that software applications (Apps) rely on. Such weaknesses can get exploited by hackers to attack Apps for data theft, remote access, or ransomware deployment (see Figure~\ref{fig:threat-model}).   

SSC has recently become a primary attack surface~\cite{software-supply-chain-vulnerabilities}. In 2025 alone, there is a 73\% increase in malicious open-source packages or Libs, revealing a troubling acceleration of SSC risks as these Libs may ultimately get used by lots of downstream Apps~\cite{reversing-labs}. Cyble's data---based on attacks claimed by threat actors on dark web data leak sites---shows that SSC attacks surged in October 2025 (i.e., 41 attacks), setting a new record more than 30\% higher than the previous peak in April~\cite{surge-attack}. 
High-profile attacks like SolarWinds~\cite{solarwind}, Log4j~\cite{log4j-2}, and MOVEit~\cite{moveit} highlighted how a single compromised library dependency can get exploited to affect
 thousands of software systems built on top of it. SSC attacks have become increasingly severe in terms of their breadth, depth, and overall impact. 

\emph{Motivation:} To mitigate supply chain attacks, people created tools to detect vulnerable Lib dependencies ~\cite{dependencychecker,snyk} or vulnerable API usage~\cite{rahaman2019cryptoguard,findsecbugs,ponta2020detection} in Apps, and to suggest repairs for security hardening~\cite{dependabot,Zhang2022,Fu2022}. However, none of these tools demonstrate how Lib vulnerabilities enable practical exploits that target downstream Apps. Recent work ~\cite{zhang2022automatic,Kabir2022} shows that given vulnerability reports, developers often require concrete and executable code to demonstrate proof-of-concept exploits, in order to assess the security risks and prioritize bug fixes.
We use \textbf{Proof-of-vulnerability (PoV) test} to refer to such demo code that exploits vulnerabilities in Libs to launch attacks to Apps. 

\emph{Problem Statement:} Given a vulnerable Lib and an App calling vulnerable APIs of that Lib, we aim to generate PoV test(s) for Apps to demonstrate 
the malicious program behaviors enabled by the vulnerable Lib version. 
There are only three tools to solve this problem~\cite{iannone2021toward,kang2022test,zhang2025can}. Experimentally, they all fail to generate PoV tests in most cases~\cite{zhang2025can}. 

\begin{figure}
   \centering
    \includegraphics[width=\linewidth]{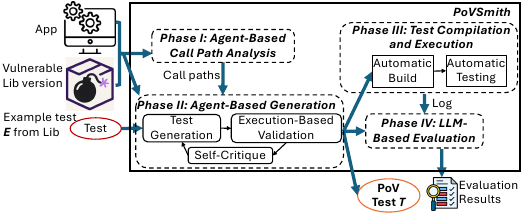}
    \vspace{-2.6em}
    \caption{\tool has four phases}\label{fig:overview}
    \vspace{-2.em}
\end{figure}

\emph{Our Solution:} To bridge the gap between developers' need and current tool limitation, we introduce \tool, an approach to generate PoV tests as JUnit tests customized for Java-based Apps. As shown in Figure~\ref{fig:overview}, \tool has four phases: call path analysis, PoV test generation, test execution, and test evaluation. 

Given a vulnerable Lib version and an App using that Lib, 
Phase I prompts an AI coding agent---Codex~\cite{codex}---to search the program context of App, for \emph{a)} public methods calling the vulnerable Lib API(s) and \emph{b)} the complete call path of each found method.
With call paths and a user-provided exemplar test, Phase II prompts Codex to iteratively generate a PoV test $T$, assess the test based on its build/execution status, diagnose the root cause for any encountered unexpected build/test issue, and refine $T$ accordingly. 
Phase III takes $T$ from Phase II, conducts automatic build, and further executes $T$ if it builds successfully. 
Phase IV takes in the build/execution logs output by Phase III together with $T$, prompting a large language model (LLM)---GPT---to evaluate whether the generated test successfully triggers the vulnerability. Phase IV outputs the evaluation result, which is sent together with $T$ for developer review. 

For evaluation, we applied \tool to a dataset of \totalApps $\langle App, Lib \rangle$ Java program pairs, where each App depends on a vulnerable Lib and the vulnerabilities correspond to \numOfCve CVE entries. 
In Phase I, \tool revealed \totalentry unique app-level entry points (i.e., public methods) calling vulnerable APIs; \totaltest of these methods were correctly found, together with their call paths properly identified. With such call path information, \tool generated \totaltest tests,  \totalcompile of which compiled smoothly and 
\totalsuccess tests successfully triggered vulnerabilities.  
Phase III correctly labeled \totalcorrectlabel tests.
 We also experimented with two alternative AI coding agents (i.e., Gemini Code Assist~\cite{gemini-code-assist} and 
 Mistral Vibe~\cite{mistral-vibe}), 
{finding Codex to work most effectively}.
Finally, we empirically compared \tool with a state-of-the-art LLM-based approach~\cite{zhang2025can}, {finding \tool to work much better by producing a lot more vulnerability-triggering tests.}

To sum up, this paper makes the following contributions: 
\begin{enumerate}
    \item \textbf{New methodology.} We invented an agent-based approach \sysname. It detects public methods in Apps that (in)directly call vulnerable APIs in Libs, iteratively generates PoV tests for each method, evaluates tests via automatic build and testing, and assesses test quality via LLM-based self-critique.
     \item \textbf{New characterization of agent and LLM capabilities.} We systematically evaluated \tool's effectiveness in call path analysis, iterative PoV test generation, and test assessment. We characterized its effectiveness when using different coding agents. We also empirically compared \tool with a state-of-the-art tool~\cite{zhang2025can}.
    \item \textbf{New dataset.} We created a dataset of \totalCorrectCallPath verified call paths, showing how vulnerable APIs of Libs can influence runtime behaviors of Apps. They belong to \totalApps $\langle App, Lib\rangle$ pairs and present the supply chain risks related to 20 CVEs.
   \item \textbf{New software security contributions.} Some of the tests output by Codex demonstrate vulnerabilities never reported before, so we filed six pull requests (PRs) and six CVE requests
   to explain those vulnerabilities and PoV demos. So far, we have not received feedback on them.
\end{enumerate} 
In the following sections, we will introduce the background knowledge (Section~\ref{sec:background}),  describe our approach (Section~\ref{sec:approach}), and discuss our experiments as well as results (Section~\ref{sec:experiment}).

\section{Background and Motivation}
\label{sec:background}
This section introduces the threat model of supply chain attacks, and AI coding agents.

{\textbf{The Threat Model of Supply Chain Attacks.}}
As illustrated in Figure~\ref{fig:threat-model}, 
our threat model 
assumes that an adversary gets to know vulnerabilities in Libs through (1) contributing flawed code to open-source projects, (2) inspecting library implementation, or (3) consulting publicly available vulnerability databases like CVE and NVD~\cite{zhang2025can}. 
The adversary's goal is 
to exploit those vulnerabilities to compromise downstream Apps.
 An attack is considered successful if the adversary's malicious input propagates from the App interface (e.g., a public method call) to Libs, triggering vulnerabilities there and causing Apps to behave abnormally. Our goal is to generate PoV tests for Apps to simulate such attacks. 

{\textbf{AI Coding Agents.}} 
Agents like Codex~\cite{codex}, Gemini Code Assist~\cite{gemini-code-assist}, and Mistral Vibe~\cite{mistral-vibe} 
act as coding partners that can be instructed via natural language to comprehend programs, generate boilerplate code, and debug as well as refactor code~\cite{agentic-coding}. 
To maintain existing code bases with the help of such agents, developers can integrate agents into the code bases through IDE extension, a Command Line Interface (CLI), or a conversational app. 
Once integrated, agents work by reading existing files, running tests or linters, and proposing changes or full features across multiple files. 
AI coding agents seem promising to help with our test generation task, although no prior work explores their usage in this context. 
This fact motivated us to  investigate using these agents when designing our automatic approach of PoV test generation.

\section{Our Approach: \tool}
\label{sec:approach}
To automatically generate PoV tests for a given $\langle App, Lib\rangle$ pair, we need to tackle four technical challenges:

\begin{itemize}
    \item[\textbf{C1.}] How can we identify the attack surface or entry points in the App to exploit the Lib vulnerability?
    \item[\textbf{C2.}] How can we create any test to effectively demonstrate the exploit logic?
    \item[\textbf{C3.}] How can we flexibly adapt the test as needed, to satisfy implicit data/control constraints posed by the App's context?
    \item[\textbf{C4.}] How can we automatically assess the quality of tests?
\end{itemize} 
We created a new agent-based approach \tool. As shown in Figure~\ref{fig:overview}, \tool has four phases: Phase I overcomes challenge C1, Phase II addresses C2--C3, while Phases III and IV handle C4. 

\textbf{Phase I: Agent-Based Call Path Analysis}.
\tool takes as input (1) a vulnerable Lib and (2) an App dependent on Lib. It applies Codex for call path analysis, to reveal 
App's public methods that (in)directly invoke vulnerable Lib API(s) and their call paths.

\textbf{Phase II: Agent-Based Generation}
\tool prompts Codex to iteratively generate, validate, and self-criticize JUnit tests, to flexibly adapt any generated test to the program context of App.

\textbf{Phase III: Test Compilation and Execution.} \tool takes in the generated test $\boldsymbol{T}$, and produces a log $\boldsymbol{L}$ to show the automatic build and/or execution status of that test. 

\textbf{Phase IV: LLM-Based Evaluation.} \tool takes in $\boldsymbol{T}$ and $\boldsymbol{L}$, to decide whether the test triggers the vulnerability successfully. 

This section introduces each phase in detail (Sections~\ref{sec:phase-1}--\ref{sec:phase-4}).

\subsection{Phase I: Agent-Based Call Path Analysis}\label{sec:phase-1}
When an App depends on a vulnerable Lib, 
the vulnerability can get exploited if (1) the App invokes vulnerable Lib API(s) and (2) the invoked API(s) are reachable from the App's public methods, which methods are directly callable by hackers to inject malicious inputs. Therefore, to identify the attack surface in App, we treat vulnerable Lib APIs as sink methods, and conduct call path analysis within App to identify source methods---public methods calling those APIs directly or indirectly.

There are program analysis tools usable for call graph analysis, such as WALA~\cite{wala}, Soot~\cite{soot}, and CodeQL~\cite{codeql}. However, most of these tools (e.g., WALA and Soot) suffer from two limitations. First, they struggle with modern JDKs (JDK 9+). When being applied to Java programs that use new language features (e.g., switch expression), these tools cannot properly handle those features or smoothly analyze software as expected. Second, they focus on bytecode analysis, while source code analysis is more desirable as we can easily map the analysis results to source code lines. 

We explored to use the state-of-the-art tool, CodeQL~\cite{codeql}, without much success. 
CodeQL is a domain-specific language defined for source code analysis. With this language, users can define CodeQL queries in a declarative way (similar to defining SQL queries) to retrieve facts from code bases, and reason about program properties (e.g., matching vulnerability patterns) with those facts. 
Based on our experience, however, CodeQL does not provide satisfactory results for two reasons. First, declarative custom queries are hard to debug: the debugging information is rare and there is no intermediate status to inspect. 
 Second, when analyzing some programs, CodeQL produces many false positives and false negatives. 

\begin{figure}
    \centering
    \includegraphics[width=\linewidth]{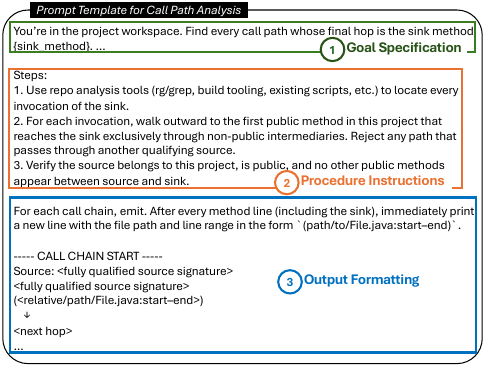}
    \caption{A simplified version of the prompt template we used to enumerate  source-to-sink call chains in Apps}
    \label{fig:callpath-prompt}  
\end{figure}

After trying existing tools
 and identifying their limitations in call path analysis, we decided to explore using an AI coding agent: GPT-5.2-Codex~\cite{gpt-52-codex}. 
 Given the software implementation of App, we wrote a script to integrate Codex into the code base via CLI, so that Codex can easily access the program context. 
 Next, we defined a prompt template to derive project-specific, vulnerability-specific, and context-aware prompts that instruct Codex to conduct call path analysis. As shown in Figure~\ref{fig:callpath-prompt}, the template consists of three parts.

 \textbf{\textcircled{1} Goal Specification} defines the goal of finding call paths ending at the given sink---a vulnerable API. The sink is presented with its fully qualified name: \codefont{ClassName.methodName(parameterlist)}. 
 
\textbf{\textcircled{2} Procedure Instructions} define a three-step procedure. Step 1 locates all invocations of the sink method. For each invocation, Step 2 identifies the call path starting from a public method, ending at the invocation, and covering solely non-public methods as intermediate nodes between the two ending points. In this way, we can reveal the attack surface in App. Notice that if a public method $A$ calls another public method $B$ that directly calls a vulnerable API, we do not consider $A$ as the attack surface. This is because any exploit of $A$ can get redirected to $B$ through method calls.  When lots of methods call $B$ (in)directly, 
there can be many redundant call paths revealed, significantly increasing the workload of PoV test generation although the extra value of traversing those redundant paths is little. Our research explores the feasibility of exploiting Lib vulnerabilities in Apps, so we focus on the minimal attack surface in Apps.  
Step 3 verifies each recognized source method, ensuring them to satisfy all our  requirements. 

\textbf{\textcircled{3} Output Formatting}
 defines the output structure, to facilitate later test generation and our manual inspection. 

Given the code base of a software project, this analysis module outputs one or more text files, with each file showing all located call paths ending at a given sink method. 

\begin{figure}
    \centering
    \includegraphics[width=\linewidth]{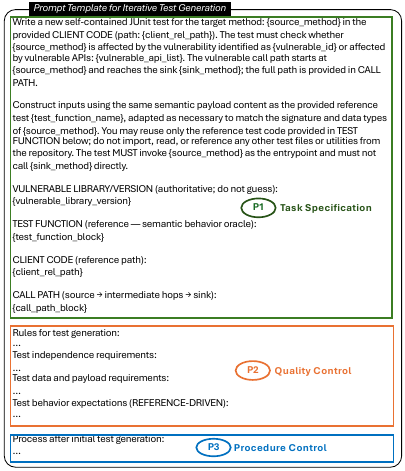}
    \caption{The template used for iterative PoV test generation}
    \Description{s}
    \label{fig:gen-prompt}
\end{figure}

\subsection{Phase II: Agent-Based Test Generation}
\label{sec:phase-2}
To generate a PoV test for each revealed call path, we explored to use Codex again via CLI due to its claimed capabilities of autonomous reading, generating, editing, and debugging code. 
Figure~\ref{fig:gen-prompt} illustrates our template design for an iterative process of PoV test generation. As shown in the figure, there are three major parts in the template.

\subsubsection{P1: Task Specification}
Our hypothesis is \emph{when we provide essential information about (1) the program context of App, (2) the vulnerability in Lib, and (3) an exemplar test $\boldsymbol{E}$ from Lib to show an effective exploit of the vulnerability, Codex can mimic that test and similarly generate a PoV test $\boldsymbol{T}$ for App}. Our approach design includes the exemplar Lib test for two reasons. First,
developers often create such tests after fixing vulnerabilities in Libs, 
as these tests enable differential testing and present distinct behaviors between vulnerable and patched Lib versions.
Such tests are often available~\cite{kang2022test}. 
Second, prior work~\cite{zhang2025can} shows that 
offering exemplar tests can significantly improve LLMs' capability of PoV test generation. 

To verify our hypothesis, we designed the template to customize the task specification for each call path using nine inputs:

\begin{itemize}
    \item[\textbf{(i)}]  \codefont{sink\_method}: a vulnerable API called by App.
    \item[\textbf{(ii)}] \codefont{source\_method}: a public method $M$ located in Phase I that calls the vulnerable API.
    \item[\textbf{(iii)}] \codefont{client\_rel\_path}: the relative path of the Java file defining $M$. 
    \item[\textbf{(iv)}] \codefont{CALL\_PATH}: the call path located for $M$ by Phase I.
    \item[\textbf{(v)}]  \codefont{vulnerable\_id}: the assigned unique identifier (i.e., CVE or JIRA entry ID) that corresponds to a website to explain details of the vulnerability, such as issue description, CVSS score, and references to solutions or tools~\cite{cve-2018-1000632}. 
    \item[\textbf{(vi)}]  \codefont{vulnerable\_api\_list}: the list of known vulnerable APIs in Lib.
    \item[\textbf{(vii)}] \codefont{VULNERABLE LIBRARY/VERSION}: the versions of Lib that are affected by the vulnerability. 
    \item[\textbf{(viii)}] \codefont{test\_function\_name}: the name of an exemplar test from Lib, which shows a potential way of exploiting the vulnerability by directing calling vulnerable API in Lib.
    \item[\textbf{(ix)}] \codefont{TEST FUNCTION}: implementation of the exemplar test.
\end{itemize}
Inputs (i)--(iv) describe the call path output by Phase I; (v)--(vii) characterize the vulnerability; and (viii)---(ix) depict the test example.

\subsubsection{P2: Quality Control}
\label{sec:quality-control}
To help Codex generate high-quality PoV tests with the best effort, we defined various rules and requirements/expectations for it to follow or satisfy.
These constraints were defined based on our initial experience of tentatively applying Codex to about 10 projects.
{\textbf{Rules for test generation.}} We defined 11 rules to confine the generation procedure from 5 angles. 
First, \emph{mocking}. Do not mock 
any class or method mentioned in the call path; mock program entities (e.g., Java classes) if they are not mentioned in the call path, but are necessary for a successful test compilation. 
Second, \emph{test configuration}. Use the test framework already configured in the project; skip 
comparing checksum/hash/signature, as they are designed to ensure reproducible builds but irrelevant to PoV testing.
Third, \emph{test structure.} Use the Arrange-Act-Assert pattern~\cite{aaa} to write tests. Namely, each test should first arrange inputs to set up the test case, then act on the target behavior (i.e, call the source method), and finally assert expected outcomes by verifying runtime behavior. Fourth, \emph{bug fixing.} Apply minor fixes as needed to make the test compilable and runnable. 
Fifth, \emph{commenting.} Add comments to describe (1) the observed behavior corresponding to the exemplar test, (2) the expected behavior when vulnerability is present, and (3) the vulnerable library as well as version mentioned in (vii). 

The first four angles guide \tool to generate good PoV tests that are likely to compile and run; the fifth one ensures the tests to be human-readable, to facilitate later manual validation.

{\textbf{Test independence requirements.}}
We defined three requirements, asking the agent to generate new tests without referring to, depending on, or modifying any existing test in App.  
{\textbf{Test data and payload requirements.}}
There are three requirements, instructing Codex to 
construct all required payloads and inputs  programmatically in the test. For instance, 
if a test requires to access an external file, Codex is supposed to create that file at runtime using a temporary directory or in-memory buffer. 

{\textbf{Test behavior expectations (REFERENCE-DRIVEN).}} 
We expressed six expectations for alignment between the new test $T$ and exemplar test $E$ in terms of their inputs, runtime behaviors, and outcomes. Specifically, $T$ should use the same semantic payload content as $E$, 
but can adapt the content as needed to provide input(s) to \codefont{source\_method}.
Second, $T$ should present runtime behaviors as close as possible to that of $E$. Namely, if $E$ produces an exception chain when the vulnerability is present, $T$ should also produce a chain, with at least the type of the first exception in the sequence to match.
However, if $E$ does not produce any exception given the vulnerability existence, $T$ should produce an exception in this circumstance, but produce a different exception when the vulnerability is absent.
Here, the test outcome must be determined by actual program execution, not by mocked or speculated execution. 

\subsubsection{P3: Procedure Control}\label{sec:procedure-control} We defined four steps to prescribe the execution-validation-generation loop, after the initial test generation. 
Step 0 determines the correct build root and test runner, to prepare for automatic build and testing.
Step 1 instructs the agent to automatically build or run the test $T$, and to observe the procedure.
Step 2 constructs a self-critique evaluation to decide whether $T$ triggers the vulnerability as expected; if not, it revises $T$ for a bug fix.  
Step 3 repeats Steps 1-2 for up to 5 times, until the test meets all expectations or until Codex finishes all trials. 

\subsection{Phase III: Test Compilation and Execution}
\label{sec:phase-3}
One challenge of applying coding agents is the hallucination issue: agents may commit mistakes in Phases I \& II, and produce wrong but seemingly correct tests. To identify such issues and avoid misleading developers with wrong data, this phase automatically compiles and runs each generated test. 

Specifically, we wrote a script to take as input (1) source code of the App, (2) the generated test $T$, and (3) project-specific build configuration. 
Based on the build configuration, our script invokes the corresponding build system (i.e., Maven or Gradle), builds $T$, and runs $T$ when it compiles successfully. Our script outputs a log file $L$, to document all compilation results, runtime logs, stack traces, and test outcomes.
Such information can help Phase IV later identify the hallucination issues concerning software build/execution status, and mitigate their potential impact on developers' decision making.

\begin{figure}
    \centering    \includegraphics[width=.75\linewidth]{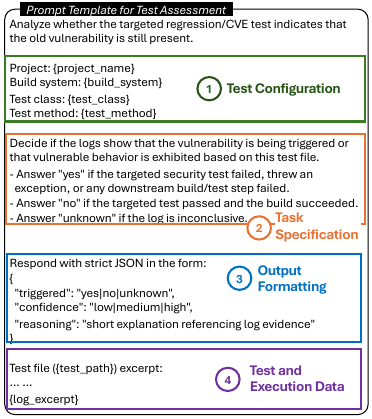}
    \caption{The prompt template we used to assess test quality}
    \label{fig:gpt-prompt}
\end{figure}

\subsection{Phase IV: LLM-Based Evaluation}
\label{sec:phase-4}

To mitigate the impact of agents' hallucination issues, this phase uses GPT-5.1~\cite{gpt-51} to determine whether $T$ successfully demonstrates a proof of vulnerability.

As shown in Figure~\ref{fig:gpt-prompt}, 
we defined a prompt template that queries GPT with the log file $L$ and test file $T$ for each test-generation task.
The template consists of four major parts:

\textbf{\textcircled{1} Test Configuration} specifies the project name, build system, test class, and the test method.

\textbf{\textcircled{2} Task Specification} 
requests GPT to examine 
both $T$ and $L$, to decide whether $T$ triggers the vulnerability as expected. 

\textbf{\textcircled{3} Output Formatting} prescribes GPT's output format, including (1) the triggering judgment, (2) GPT's confidence in that judgment, and (3) its explanation. 
Such information assists us to later manually inspect all sorts of software artifacts produced in the automatic procedure, to investigate \emph{(a)} agents' hallucination issues in test generation, \emph{(b)} LLM's hallucination issues in automatic judgment, and \emph{(c)} \tool's overall capabilities in test suggestion.

\textbf{\textcircled{4} Test and Execution Data} presents the detail of $T$ and $L$.

\noindent
\tool composes $L$ using two complementary files: a \codefont{.txt} file showing the human-readable log and a \codefont{.jsonl} file showing a machine-readable execution summary.
In particular, 
the text file contains the lengthy raw log, including stack traces of failed tests and any intermediate steps documented in the test procedure.
The JSON file was generated by our in-house script in \tool, capturing only the essential data like test pass or the first line of the initial stack trace triggered by a chain of build/test failures. 

When a test gets executed, both Maven and Gradle systems produce 
a \codefont{.txt} log file and an \codefont{.xml} log.
With our initial investigation, we made two observations. First, the two log files are complementary: although their information often overlaps significantly, there is data sometimes only mentioned by one file but not by the other. 
Second, the two log files are so lengthy that sending them directly to GPT can considerably mislead the model to wrongly assess tests.

Inspired by these observations, 
we designed \tool to query GPT with the raw \codefont{.txt} log file and a simplified version of the \codefont{.xml} log, i.e., \codefont{summary.jsonl}. 
Namely, 
\tool extracts a test-relevant excerpt from each XML file via pattern matching, using the class name and optionally method name as the keyword(s). 
Next, it converts the excerpt into a customized structural representation---a JSON format---to facilitate LLM processing.
For any plain Java project that is not built with  Gradle or Maven, \tool creates the JSON file by checking the exit code. If exit code is 0, the file documents the test to pass; otherwise, it records a \codefont{CommandFailure}.
\section{Evaluation}
\label{sec:experiment}
In our experiments, we explored five research questions (RQs): 

\begin{itemize}
    \item \textbf{RQ1:} How effectively does \tool analyze call paths?
    \item \textbf{RQ2:} How effectively does \tool generate tests?
    \item \textbf{RQ3:} How effectively does \tool assess test quality?
    \item \textbf{RQ4:} How effective does \tool work when it uses different coding agents?
    \item \textbf{RQ5:} How well does \tool compare with prior work?
\end{itemize}
This section presents the experiment dataset we used for evaluation (Section~\ref{sec:dataset}), 
our experiment settings for RQs (Section~\ref{sec:setup}), 
and the experimental results (Sections~\ref{sec:rq1}--\ref{sec:rq5}).

\subsection{Dataset Curation}
\label{sec:dataset}
We reused the dataset created by prior work~\cite{zhang2025can}. Among the 49 $\langle App, Lib\rangle$ Java program pairs in the original dataset, we managed to include \totalApps Apps. We could not include the remaining Apps because (1) we could not compile 15 Apps, and (2) one app depends on a patched instead of vulnerable version of Lib. 
Two possible reasons may explain the compilation failures. First, 
the App versions we downloaded are different from
the App versions used by prior work. Second, 
some required project dependencies are not retrievable. 
As a result, our dataset consists of \totalApps Java-based $\langle App, Lib\rangle$ program pairs: covering 29 Maven projects, 3 Gradle projects, and 1 plain Java project without Maven or Gradle. 
It corresponds to \numOfCve vulnerability entries.
The vulnerable libraries cover various domains like data processing (e.g., Apache Commons Codec~\cite{apache-commons-codec}) and security (e.g., Spring Security~\cite{spring-security}). 
These entries 
can be exploited to realize four  kinds of attacks: \textbf{denial of service ({DoS})}, \textbf{directory traversal ({DT})}, \textbf{remote code execution (\textbf{RCE})}, and \textbf{others ({OTH})}.

We built all Apps with a computer equipped with an Intel Core i9-9900 CPU running at 3.1 GHz,
32 GB RAM,
2.25 TB storage, and an  Intel UHD Graphics 630 (CoffeeLake-S GT2) GPU.


\subsection{Experiment Settings} \label{sec:setup}
This section introduces our experiment design. 
 RQ1--RQ3 examine the effectiveness of Phases I--III in \tool (Sections \ref{sec:i-rq1}--\ref{sec:i-rq23}). 
RQ4 is an ablation study of \tool (Section~\ref{sec:i-rq4}). We empirically compared \tool with prior work for RQ5 (Sections~\ref{sec:i-rq5}). 

\subsubsection{Investigation of RQ1}\label{sec:i-rq1}
We applied \tool to the dataset of Apps, and got \totalCallPath public methods reported as the application-level entry points, corresponding to \totalentry unique methods. 
Two authors separately inspected (1) each reported method, (2) the revealed call path, and (3) the source code, to decide whether the call paths are correctly reported. They then cross-checked each other's results,
discussing and resolving any differences until reaching a consensus. 

\subsubsection{Investigation of RQ2 an RQ3}
\label{sec:i-rq23}

When generating prompts for test generation, we started with the  correct call paths derived from Phase I. 
We sampled a call path for each of the 
correctly located unique methods, to define distinct test generation tasks. 
If a method has alternative call paths identified, we sampled the longest path, breaking ties by randomly choosing among those of equal maximum length. Our insight is that when call paths are longer (e.g., greater than 1), the generation tasks often demand inter-procedural program analysis and thus become more challenging for any tool to create project-specific tests.
Consequently, in total, we have \totaltest generation tasks for \tool to perform.

Given those tasks, 
\tool output generated tests, their build/ execution logs, and GPT's assessment of the test quality. We manually inspected all outputs together with the corresponding prompts to decide (1) whether each test successfully demonstrates PoV, and (2) whether the GPT's assessment is correct. 
A test $T$ is considered successful 
it satisfies three conditions:

(C1) $T$ reuses the inputs $E$ sends to the vulnerable API (i.e., payload), to construct inputs sent to $M$.

(C2) $T$ defines the same or semantically relevant expected behaviors as $E$ for the vulnerable program. 

(C3) The execution log $L$ confirms that $T$ presents the dynamic behaviors of the vulnerable program.

\begin{figure}
    \centering
    \includegraphics[width=\linewidth]{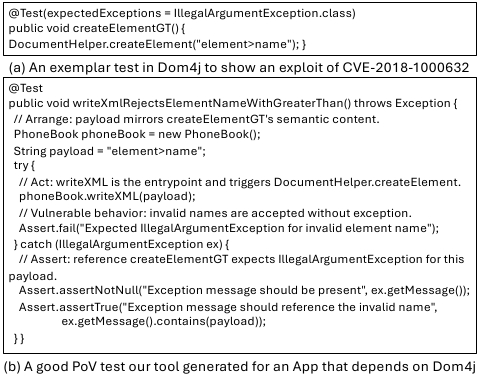}
    \caption{One PoV test that \tool successfully generated}
    \label{fig:good-test}
\end{figure}

\begin{figure}
    \centering
    \includegraphics[width=.85\linewidth]{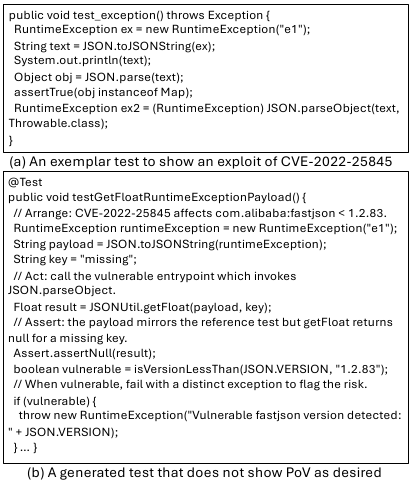}
    \caption{One generated test that fails to demonstrate PoV}
    \label{fig:bad-test}
\end{figure}

For instance, Figure~\ref{fig:good-test} shows a PoV test that \tool successfully generated. The exemplar test from Dom4j~\cite{dom4j} (see Figure~\ref{fig:good-test}(a)) demonstrates that a vulnerable version of Dom4j silently accepts an illegal input \codefont{``element>name''}, allowing attackers to tamper with XML documents through XML injection. However, if this test executes with a patched version of Dom4j, an IllegalArgumentException is expected to be thrown to reject the malicious input.
Given this test, \tool successfully generated a PoV test for method \codefont{PhoneBook.writeXML(...)} in the App tcpser4j~\cite{good-test-tcpser4j}, which method directly calls the vulnerable API \codefont{DocumentHelper.createElement(...)}. 
As shown in Figure~\ref{fig:good-test}(b), this test reuses the payload; it similarly defines the program logic as $E$ but has a clearer differentiation between vulnerable and patched program behaviors.
After executing $T$ with the App on top of a vulnerable version of Dom4j, we inspected the log and found an AssertionError \codefont{``Expected IllegalArgumentException for invalid element name''}, to signal a successful vulnerability triggering.

As another example, Figure~\ref{fig:bad-test} shows a generated test that fails to demonstrate PoV. The exemplar test from Fastjson~\cite{fastjson} (see Figure~\ref{fig:bad-test}(a)) shows that the vulnerable API \codefont{JSON.parse(...)} silently accepts an illegal input, and leverages the flawed handling of \codefont{Throwable} classes to bypass autoType restrictions, allowing attackers to instantiate arbitrary classes on the target classpath and achieve Remote Code Execution (RCE).
Given this test, \tool generated a test for method \codefont{JSONUtil.getFloat(...)} in the App commerce~\cite{commerce}, which method directly calls the vulnerable API.  
As shown in Figure~\ref{fig:bad-test} (b), this test reuses the payload to similarly call \codefont{JSONUtil.getFloat(...)}. However, instead of examining the method's return value for vulnerability detection, $T$ checks if the current Lib version is less than \codefont{1.2.83}. 
The log $L$ shows \codefont{``java.lang.RuntimeException: Vulnerable fastjson version detected''}, but it does not confirm that $T$ 
achieves RCE.
Thus, among the three conditions mentioned above, $T$ satisfies C1, but dissatisfies C2--C3; it does not demonstrate PoV as desired.

To quantitatively evaluate \tool's capability of test generation, we defined four metrics:

\textbf{Test Compilability (C)} counts the number of generated tests that compile successfully.

\textbf{Compilability Rate (CR)} measure the percentage of generated tests that compile successfully.

\textbf{PoV Demonstration (D)} counts the number of generated tests that compile and trigger vulnerabilities successfully.

\textbf{Demontration Rate (DR)} measures the percentage of generated tests that demonstrate PoV successfully.

Based on our experience, the tests generated by \tool often look relevant to exemplar tests and reuse payloads in delicate ways, which made it very challenging for us to identify tests that do not trigger vulnerabilities as expected. To ensure the quality of our results, two authors separately inspected all artifacts to label the results; they sometimes even conduct step-by-step debugging to collect additional intermediate execution data when logs provide insufficient detail. They cross-checked each other's results, and involved a third author as needed to resolve the labeling differences.

\subsubsection{Investigation of RQ4}\label{sec:i-rq4} 
By default, \tool leverages GPT-5.2-Codex 
in test generation (i.e., Phase II). To explore how our agent selection affects the overall effectiveness of test generation, we also explored two alternative agents: Gemini Code Assist with Gemini-2.5-pro and Mistral Vibe 2.0 with Devestral 2.
It has been very time-consuming to manually validate all generated tests for the \totaltest tasks. Due to the time limit, we 
randomly selected one test generation task for each of the \totalApps $\langle App, Lib\rangle$ pairs, to manually validate \tool's outputs when the tool uses different agents.   

To quantitatively measure \tool's capability of test assessment, we defined one metric:

\textbf{Assessment Accuracy (A)} measures the percentage of LLM-based judgments to be correct. 

\subsubsection{Investigation of RQ5}\label{sec:i-rq5} 
Zhang et al.~\cite{zhang2025can} created an LLM-based approach of PoV test generation. As with our work, that approach takes (1) code content of App, (2) vulnerability information, and (3) exemplar tests from Lib as input, to output PoV tests for Apps. However, unlike our work, it relies on users to manually specify app-level entry points, instead of automatically revealing those methods.
It does not iteratively generate tests, neither does it adopt compilation/testing or LLM-based assessment to automate test validation.  
The researchers empirically showed that their  approach outperforms existing tools SIEGE~\cite{iannone2021toward} and Transfer~\cite{kang2022test}. 
Inspired by their work, we applied the LLM-based approach to our sample dataset, to explore how \tool compares with the state of the art. Although Zhang et al.~proposed their approach to use ChatGPT-4.0, we applied GPT-5.4 
to leverage the state-of-the-art GPT model, and to eliminate any disadvantage caused by model selection.

\subsection{Call Path Analysis Results (RQ1)}
\label{sec:rq1}

\begin{figure}
    \centering
\includegraphics[width=.7\linewidth]{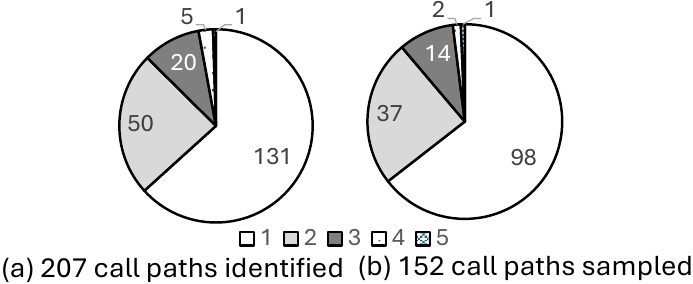}
    \caption{Distribution of call paths by length }
    \label{fig:call-paths}
\end{figure}

Among the \totalCallPath call paths revealed, \totalCorrectCallPath call paths are correct. It indicates that 
\tool works very effectively; it achieved a very high precision, i.e., 96\% (\totalCorrectCallPath/\totalCallPath), in identifying the call paths satisfying our requirements.
In particular, as shown in Figure~\ref{fig:call-paths}(a), 131 paths have length 1, meaning that the identified  methods directly call vulnerable APIs; 
the other 76 paths have longer lengths (i.e., 2--5), showing how the identified methods indirectly call APIs.
These \totalCallPath paths correspond to \totalentry unique source methods, as some paths share the same source and sink methods.

For the nine call paths that were incorrectly identified, we observed two major phenomena. First, in three cases, 
either the sink or source method was wrongly recognized, because (1) \tool was sometimes confused by overridden methods and (2) a located source method is a public method defined in an anonymous class, which method is not callable or exploitable by code outside of the App.
Second, in six cases, the locations of some reported methods on call paths got wrongly described. Namely, a method was either described to be in a wrong file or at a wrong source line. We believe  such mistakes easy to fix, if  
we apply basic static program analysis to screen \tool's results in the future. 

Additionally, we contrasted the \totalentry unique methods  against the \totalApps public methods manually labeled for these \totalApps Apps in prior work~\cite{Zhang2024how}, to examine whether \tool missed any known call paths. We found \tool to miss 8 of the \totalApps methods. Among these eight methods, one is declared protected rather than public, and another is a public method that invokes the vulnerable API indirectly via another public method; 
neither qualifies as an application-level entry point under our definition. Five of the missing methods call vulnerable APIs directly, a sixth missing method calls the vulnerable API indirectly. Therefore, \tool achieved 81\% (25/31) recall, with the F1-score as 88\%.

\noindent\begin{tabular}{|p{8.2cm}|}
	\hline
	\textbf{Finding 1 (Answer to RQ1):} \emph{\tool demonstrated great potentials in call path analysis. When detecting application-level entry points for vulnerable APIs and reporting their call paths, \tool obtained 96\% precision, 81\% recall, and 88\% F1-score.
    }
	\\
	\hline
\end{tabular}

\subsection{Test Generation Results (RQ2)}
\label{sec:rq2}
\tool performs impressively in generating PoV tests. 
This section presents and explains both (1) the quality of generated tests, and (2) the procedure of test generation.
\subsubsection{Compilation and Execution Status of Generated Tests}
We observed \totalcompile of the \totaltest tests (92\%) compile successfully; the remaining 12 tests fail compilation for 3 reasons.
First, Codex incorrectly judged the build to succeed for nine cases.
Second, it incorrectly diagnosed the root cause of compilation failures in two cases, and thus was unable to resolve failures. 
Third, it correctly found the root cause for one case (i.e., dependency resolution failure) but could not resolve that issue.

There are \totalsuccess generated tests successfully demonstrating PoV, while 68 tests failed to do so. In addition to the  compilation failures mentioned above, three reasons explain unsuccessful cases. First, 37 tests 
 presented non-vulnerable program behaviors, such as behaviors of patched versions or behaviors different from the expected behaviors of vulnerable versions.
This implies that either App developers applied local patches to address Lib vulnerabilities, or their usage of vulnerable API is not exploitable. 
For these cases, Codex could not identify the unexploitability of vulnerable API usage, but just blindly generated PoV tests with the best effort.

Second, 11 tests threw exceptions (i.e., \codefont{NoSuchMethodException}) before calling vulnerable APIs, 
meaning that the exceptions are totally irrelevant to Lib vulnerabilities. These exceptions were due to either malformed input parameters sent to method calls, or improper setups of complex data structures. 
Third, seven tests did not demonstrate PoV, as the payloads (i.e., malicious inputs) from reference tests were not used to call vulnerable APIs. 
Fourth, two tests did not call the methods-under-test as instructed.

Among these \totaltest tasks, vulnerabilities can be exploited to achieve 4 kinds of attacks: DoS (60), DT (13), RCE (30), and OTH (49).
Within these categories,
\tool successfully generated PoV tests for 30, 10, 3, and 41 tasks, respectively. Its demonstration rates (DR) per category are separately 50\%, 77\%, 10\%, and 84\%, implying our tool's better effectiveness in demonstrating DT and OTH.

\subsubsection{Test Evaluation via Codex Self-Critique}
When assessing the compilation status of generated tests, Codex was correct in 134 cases (88\%) and incorrect in 18 cases. In the latter, it either classified compilable cases as incompilable or vice versa. When evaluating the PoV demonstration status of generated tests, 
Codex was correct in 94 cases (62\%) and incorrect in 58 cases (38\%).
These phenomena imply that although Codex's feedback mechanism (execution-based validation + self-critique) is effective in most cases,
it alone is insufficient to reliably evaluate test quality.

\subsubsection{Attempts and Generated Tests}
We clustered tests based on the number of trials \tool made to output each test.
Namely, we manually inspected the execution logs of Codex during its iterative procedure. Considering  each round of ``test (re)generation $\rightarrow$ test execution $\rightarrow$agent's self-critique'' as one attempt, we manually counted the number of attempts in each log. We identified two key findings from the data. First, \emph{the majority of tests (around 75\%) were output by \tool based on multiple attempts instead of single trials}.  This implies the necessity of defining an iterative workflow for test generation, execution, and evaluation in Phase II, because single trials often led to tests that the agent were dissatisfied with.

Second, \emph{Codex does not strictly follow our procedural instructions.} In our prompt (see Section~\ref{sec:procedure-control}), we required it to retry up to five times.
However, for about 20\% of tests, Codex made 7 or more attempts, causing the entire procedure unpredictable. Two reasons may explain such observations: \emph{(a)} the agent suffers from hallucination issues when counting the number of retries; \emph{(b)} our prompt (see Section~\ref{sec:quality-control}) allows it to apply minor fixes as needed to make the test compilable and runnable, which inspires Codex to autonomously conduct additional trials for debugging.

 \subsubsection{Call Path Lengths and Generated Tests}
We clustered the \totalsuccess successfully generated tests based on call path lengths.
Among the 98 tests generated for public methods that directly call vulnerable APIs (i.e., with 1-length call paths), 
 56 tests (57\%) demonstrate PoV. Among the 54 tests generated for other methods, only 28 tests (52\%) demonstrate PoV. 
It means that when call paths are simpler, \tool is more likely to generate PoV tests successfully.

\noindent\begin{tabular}{|p{8.2cm}|}
	\hline
	\textbf{Finding 2 (Answer to RQ2):} \emph{
Phase II generated tests for \totaltest tasks; \totalcompile of the tests (93\%) compile smoothly, and \totalsuccess (55\%) successfully demonstrate PoV.
It assessed test compilation with 88\% (134/\totaltest) accuracy, and PoV demonstration with 62\% (94/\totaltest) accuracy.
}
	\\
	\hline
\end{tabular}

\subsection{Results of LLM-Based Assessment (RQ3)}
\label{sec:rq3}

Among the given tests, GPT-5.1 labeled 103 tests to successfully demonstrate PoV, and labeled 35 tests to be unsuccessful; it marked 14 cases with ``unknown'', meaning that it has insufficient information to assess. By manually inspecting these ``unknown'' cases, we realized that most of the cases are incompilable. Therefore, we interpreted the ``unknown'' cases as unsuccessful triggers diagnosed by GPT as well. With such an interpretation, we contrasted GPT's assessment with our manual assessment, and found GPT to be correct in \totalcorrectlabel cases (68\%).
Its accuracy is higher than that of Codex self-critique (62\%), probably because it assesses test quality based on the actual execution logs produced by our automation script, while Codex bases its self-critique on self-generated execution data that may be subject to hallucination. 
We manually inspected the confidence values output by GPT, to find any correlation between those values and the assessment accuracy. Unfortunately, we saw no correlation, as GPT always output ``high'' values for its judgments.

With further detail, among the 103 tests GPT labeled as vulnerability-triggering, 69 tests demonstrate PoV; among the remaining 49 tests, 34 tests fail to demonstrate PoV. Namely, when labeling vulnerability-triggering cases, GPT achieved 67\% accuracy (69/103). When signaling non-triggering cases, it achieved 69\% accuracy (34/49).
The comparison between 69\% and 67\% implies that 
GPT is slightly better at recognizing non-triggering cases. This may be because build errors have typical structures and frequently used keywords. When build errors are present in logs, GPT can easily identify them, and signal the demonstration failures.

The 68\% overall accuracy achieved by GPT is still not reliable enough for us to blindly trust its automatic assessment. One possible reason to explain the observed insufficiency is that the generated 
PoV tests have nontypical structures/logic and project-specific error messages to imply vulnerability triggering:  most tests signal vulnerability-triggering via failures, while other tests do so via passes; 
some tests align well with exemplar  tests by producing the same or semantically relevant errors/exceptions, while others signal vulnerability triggering with totally unrelated errors/exceptions. 

\noindent\begin{tabular}{|p{8.2cm}|}
	\hline
	\textbf{Finding 3 (Answer to RQ3):} \emph{GPT assessed PoV test quality with good accuracy (68\%); it worked slightly better at identifying non-triggering tests than labeling vulnerability-triggering ones.}
	\\
	\hline
\end{tabular}

\subsection{Ablation Study (RQ4)}\label{sec:rq4}

\begin{table*}
\caption{Empirical comparison between \tool, its variants, and Zhang et al.'s tool on the \totalApps-sample dataset}\label{tab:compare}
\footnotesize
    \centering
    \begin{tabular}{l |r| r| r| r| r| r| r }
    \toprule
    \multirow{2}{*}{\textbf{Tool}} & \multicolumn{2}{c|}{\textbf{Self-Critique}}& \multicolumn{4}{|c|}{\textbf{Test Quality}} &\textbf{LLM-Based Assessment (GPT-5.1)}\\
   \cline{2-8}
    & \textbf{Compile?} & \textbf{Demoed?} &   {\textbf{C}} &{\textbf{CR}} &{\textbf{D}} &{\textbf{DR}} & \textbf{A}\\ \hline
     \tool (using GPT-5.2-Codex) & \textbf{32} & \textbf{31} & 31 & 94\% & \textbf{23} &\textbf{70\%} & {73\%(24/33)}\\ 
   \sysnameg (using Gemini Code Assist with Gemini-2.5-pro)  & 21 & 13 & 23 & 70\% & 11 & 33\% & \textbf{79\% (26/33)}\\ 
   \sysnamem (using Mistral Vibe 2.0 with Devestral 2) & 29 & 18 & \textbf{32} & \textbf{97\%} & 12 & 36\% & {73\%(24/33)}\\ \hline
   Zhang et al's tool~\cite{Zhang2024how} & - & - & 16 &48\%&5&15\%&-\\
   \bottomrule
    \end{tabular}
\end{table*}
We revised \tool to use 
two alternative agents for test generation,
producing two variants of \tool. 
We contrasted the three versions of \tool in  their agents' self-critique capability, 
test quality, and LLM-based assessment (see Table~\ref{tab:compare}).
These versions commonly
output compilable tests for 23 cases, 
demonstrated PoV for 7 cases, and obtained correct LLM-based assessment for 16 cases.  

In terms of agents' self-critique, \sysnameg was the most pessimistic:  
it assessed 21 of the tests to compile successfully, 7 tests to fail compilation, and only 13 tests to demonstrate PoV. However, it did not assess the quality of five tests, failing to follow our instructions on self-critique. Meanwhile, \tool was the most optimistic: it judged 32 tests to compile successfully, 1 test to have no compilation verifiability due to wrong Java runtime version (hallucination), and 31 tests to demonstrate PoV. 
In terms of test quality, based on our manual validation, \sysnamem output the largest number of compilable tests (i.e., 32), while \sysnameg output the fewest (i.e., 23). \tool output the biggest number of tests to demonstrate PoV (i.e., 23), while the variants only demonstrated PoV in 11-12 cases. Eight of the compilable tests output by \tool could not demonstrate PoV: 
five cases demonstrated patched behaviors of programs; the other three caused runtime exceptions irrelevant to vulnerabilities  (i.e., \codefont{NoClassDefFoundError}).
\sysnameg and \sysnamem worked worse than \tool, as (1) they did not call source methods as instructed or (2) the test logic did not trigger vulnerabilities effectively.

In terms of LLM-based assessment, 
GPT-5.1 worked most effectively in assessing the tests of Gemini Code Assist, although the agent obtained the lowest demonstration rate (33\%). While Codex is based on GPT-5.2---an LLM from the same family, GPT-5.1 did not better assess Codex's tests than those output by  other agents. This may be because LLMs from the same family commonly share advantages and disadvantages. When Codex output tests with its best effort, GPT-5.1 was often positive about the test quality. 

Overall, \tool demonstrated PoV most effectively when using Codex, but least effectively when using Gemini Code Assist. This phenomenon implies that the usage of different agents can considerably impact the effectiveness of our approach. Codex outperforms the other agents in (1) generating code and (2) following our explicit instructions in prompts for quality assurance. 

\noindent\begin{tabular}{|p{8.2cm}|}
	\hline
	\textbf{Finding 4 (Answer to RQ4):} \emph{The effectiveness of \tool varied a lot when it used different agents: Codex enabled \tool to obtain the best results while Gemini Code led to the worst.}
	\\
	\hline
\end{tabular}

\subsection{Comparison with Prior Work (RQ5)}
\label{sec:rq5}

As shown in Table~\ref{tab:compare}, 
among the \totalApps tests output by the LLM-based approach, 16 tests compile successfully, but only 5 tests trigger vulnerabilities successfully. The majority of generated tests (i.e., 18) do not compile, mainly due to their usage of unknown packages/classes/methods/fields. 
Three reasons can explain these compilation failures. First, the original prompts in the LLM-based approach provide insufficient program context, causing GPT to output tests that do not fit into the context.
Second, the LLM-based tool does not iterativey generate tests, neither does it adopt compilation/testing or self-critique to guide test refinement. 
Third, we did not manually apply minor fixes to address compilation issues, as what Zhang et al. did.

Among the 16 compilable tests, 11 tests failed to demonstrate PoV for 3 reasons. First, four tests did not call the specified app-level entry points. Second, two tests failed before reaching the call sites of vulnerable APIs. Third, five tests did not present vulnerability-triggering program behaviors (e.g., Denial of Service). 

\noindent\begin{tabular}{|p{8.2cm}|}
	\hline
	\textbf{Finding 5 (Answer to RQ5):} \emph{\tool outperforms the state-of-the-art LLM-based approach. It successfully demonstrated PoV in 22 cases, while LLM-based approach achieved that  in only 5 cases. }
	\\
	\hline
\end{tabular}

\section{Related Work}
The work related our research includes 
automatic vulnerability detection, 
automatic vulnerability repair, 
and security test generation.

\subsection{Automatic Vulnerability Detection}
Tools were created to detect vulnerabilities in software~\cite{dependabot,dependencychecker,snyk,ponta2020detection,rahaman2019cryptoguard,zhang2022automatic,findsecbugs,kruger2017cognicrypt,du2024generalization,Guo2024Outside,Hu2023llm,Liu2023Software,Omar2023deep,Purba2023software,Sun2024gptscan,Yang2025dlap,Zhang:2025aa}.

Specifically, dependency checkers~\cite{dependabot,dependencychecker,snyk} scan software for vulnerable dependencies, and suggest secure library versions for replacement. Rule-based cryptographic API misuse detectors~\cite{rahaman2019cryptoguard,zhang2022automatic,findsecbugs,kruger2017cognicrypt} 
statically analyze programs to locate vulnerable API usage in software. 
EclipseSteady~\cite{ponta2020detection} analyzes patches applied to Lib versions, to locate Java methods changed to fix vulnerabilities and treat those methods as vulnerable APIs. It then applies static and dynamic analysis to decide whether these methods are reachable from any downstream projects, to report Apps that actually execute or can potentially execute vulnerable APIs.
All these tools have vulnerability patterns clearly defined (explicit or implicit), and implement deterministic algorithms to find vulnerabilities. 

Recently, researchers explored LLM usage for vulnerability detection. For instance, Purba et al.~\cite{Purba2023software} applied four well-known LLMs to two public datasets, to assess LLM performance in detecting software vulnerabilities. 
Guo et al.~\cite{Guo2024Outside} and Omar et al.~\cite{Omar2023deep} fine-tuned LLMs with datasets of vulnerable code.
Hu el al.~\cite{Hu2023llm} introduced an adversarial framework GPTLENS that makes LLM to play dual roles, i.e., AUDITOR and CRITIC, respectively.
Du et al.~\cite{du2024generalization} integrated multi-task learning with LLMs to effectively mine deep-seated vulnerability features.
Liu et al.~\cite{Liu2023Software} combined code retrieval with GPT-based code analysis, to leverage LLM's contextual learning ability. 
Sun et al.~\cite{Sun2024gptscan} combined GPT with static analysis.
Yang et al.~\cite{Yang2025dlap} combined deep learning with LLM. 

Our research does not detect security vulnerabilities. For any App found to be dependent on a vulnerable Lib, \tool locates app-level entry points to those vulnerable APIs, generates PoV tests to demonstrate the potential exploits of vulnerabilities, 
and motivates developers to take vulnerability reports more seriously. 

\subsection{Automatic Vulnerability Repair (AVR)}
Approaches were proposed to correct programs or repair vulnerabilities~\cite{Ma2017,Martinez2018,Liu2019,Zhang2022,Fu2022,Chen2023,Chi2023,Ahmad2024on,Wu2023how,Jin2023inferfix,Le2024a,Li2024exploring,Zhang2024evaluating,Wei2023copiloting,Yin2024,Zhao2024enhancing,zhao2024repair,Wang2024software}.
For instance, VuRLE~\cite{Ma2017} and SEADER~\cite{Zhang2022} learn vulnerability-repair patterns from $\langle insecure, secure\rangle$ code examples, which patterns were then used to detect vulnerabilities and suggest repairs. Search-based program repair tools~\cite{Martinez2018,Liu2019}  
adopt spectrum-based fault localization to find bug locations; they leverage templates automatically mined from existing code bases or prior bug fixes to generate candidate patches, and validate each patch via testing. 

Some researchers proposed learning-based AVR. 
For instance, 
based on the intuition  that the knowledge learned from bug fixes can be transferred to fixing vulnerabilities, 
Chen et al~\cite{Chen2023} trained
VRepair on a large bug fix corpus, and then tuned it on a vulnerability fix dataset. 
VulRepair~\cite{Fu2022} and SeqTrans~\cite{Chi2023} were sequence-to-sequence models, trained on historical vulnerability fixes to suggest fixes for given vulnerable code.
Ahmad et al.~\cite{Ahmad2024on}, Le et al.~\cite{Le2024a}, Pearce et al.~\cite{Pearce2023examining} separately explored the LLM usage to repair security bugs in hardware, Javascript programs, and synthetic C code. 
They all reported promising results by off-the-shelf LLMs, while highlighting challenges in generating functionally correct code. 

To further improve LLM-based AVR, Wu et al.~\cite{Wu2023how}, Li et al.~\cite{Li2024exploring}, and Zhao et al.~\cite{zhao2024repair} fine-tuned language models with automatic program repair (APR) data. 
InferFix~\cite{Jin2023inferfix} and ThinkRepair~\cite{Yin2024} both combine information retrieval with LLM-based repair generators.
Repilot~\cite{Wei2023copiloting} synergistically synthesizes a candidate patch through the interaction between an LLM and a completion engine.
DRCodePilot~\cite{Zhao2024enhancing} generates initial patches grounded in design rationale (i.e., developers' planned solutions and a set of underlying reasons before they patch code); it then refines patches based on the feedback of reference patches and identifier recommendations.

Our research does not repair vulnerabilities, but it is relevant. 
By presenting differential program behaviors, \tool's tests can help AVR decide whether a repair removes the located vulnerability. 

\subsection{Security Test Generation}
Some tools were created to generate security tests and assess vulnerability exploitability~\cite{Cha2012,Avgerinos2014automatic,Takanen2017,Alshmrany2021,Metta2022,afl,oss-fuzz}. 

Automatic exploit generation (AEG)
typically 
combines fuzz testing (fuzzing) with symbolic execution to identify software vulnerabilities, and then exploit them automatically~\cite{Cha2012,Avgerinos2014automatic,Alshmrany2021,Metta2022}.
These tools have limited applicability in practice, as symbolic execution engines are inherently incomplete; they suffer from scalability issues like path explosion.
Coverage-based fuzzing has better applicability
~\cite{Takanen2017,afl,oss-fuzz,xu2024ckgfuzzer,Zhang2024how,Lyu2024}. Fuzzers inject invalid, malformed, or unexpected inputs into an initial seed (i.e., a program test) to reveal software defects and vulnerabilities. However, fuzzers lack the domain knowledge to explore deep paths.  

Penetration testing (Pentest) involves simulated attacks against a system to identify exploitable vulnerabilities~\cite{pentest}. Manual Pentest is tedious and time-consuming, so researchers created tools based on Reinforcement Learning~\cite{Hu2020automated,Chu2018,Chaudhary2020} and LLM~\cite{Huang2024penheal,Pratama2024,Rigaki2024out}. 
For instance, PenHeal~\cite{Huang2024penheal} has two LLM-enabled components:
one module to detect vulnerabilities, and 
one module to recommend optimal remediation strategies.
Based on our best knowledge, no Pentest tool mimics the 
supply chain attacks we study.  

Our research is different from prior work in three aspects. First, it explores an automated pipeline to generate, execute, and assess PoV tests for iterative test refinement. 
It neither requires users to specify exploitability properties as AEG does, nor conducts expensive search for malicious inputs as fuzzers. Programmers can use it  in software development, to estimate the exploitability of reported dependency vulnerabilities. In this way, programmers do not have to wait till software releases to obtain Pentest results.  
Second, it explores to use Codex to conduct call graph analysis, and reveals promising results. 
Third, it reveals many app-level entry points to trigger Lib vulnerabilities, and demonstrates the exploitability.

\section{Discussion}

Our observations may be limited to the experiment dataset, experimented LLMs, and the explored agents. To ensure the representativeness of our experiments, we 
included \totalApps $\langle App, Lib \rangle$ pairs to cover \numOfCve vulnerabilities and \totaltest unique, diverse calling contexts of vulnerable APIs for test generation: 98 of these calling contexts have methods-under-test to directly call vulnerable APIs, while the remaining 54 cases have indirect calls to those APIs.  
In the future, to further strengthen the representativeness of our experiments, we will explore more vulnerabilities, more $\langle App, Lib \rangle$ pairs, and more AI models. We will contribute more manual effort to validate results produced by different AI models on larger datasets.

Our manual test validation may be subject to human bias.
Although the LLM-based validation is very efficient and can achieve up to 68\%--79\% accuracy, the accuracy is not high enough for us to blindly trust LLM's validation results. 
To ensure validation quality while mitigating human bias,  
we spent a lot of time and effort on manual validation. 
Two authors independently validated tests and documented their assessment results. They then compared results with each other, and involved a third author to resolve any result difference. 
Additionally, we treated the LLM-based validation results as second opinions, and reexamined all relevant artifacts (i.e., code and logs) if our manual results diverged from that by GPT. 


\section{Conclusion}

Software supply chains present a primary attack surface, while defending them requires  both deep software engineering (SE) expertise and strong cybersecurity knowledge.
Our work nicely integrates SE expertise with security knowledge, helping bridge the gap between software practitioners and security researchers.
We gained two major insights in our investigation. First, generating PoV tests is 
a very challenging problem. To effectively generate such tests, we explored an approach---\tool---to guide a coding agent with (1) diverse constraints on the generation procedure and test content, (2) relevant program context, (3) exemplar test, and (4) program compilation/execution status.
Second, \tool shows impressive effectiveness in detecting app-level entry points for library vulnerable APIs, 
iteratively generating PoV tests for those entry points, and assessing test quality automatically.

In the future, we will enhance automatic validation in two ways. First, execute each test $T$ with the App on top of both vulnerable and patched Lib versions, to compare the test outcomes for any difference. 
Second, analyze tests to mark (1) wrong interpretation of CVEs in comments, and (2) misleading error messages in code.

\bibliographystyle{ACM-Reference-Format}
\bibliography{main.bib}

\end{document}